# FAIR DIVISION WITH DIVISIBLE
# AND INDIVISIBLE ITEMS

Alexander Rubchinsky

State University – Higher School of Economics and International University "Dubna

**Rubchinsky A. Fair division with divisible and indivisible items:** Working paper WP7/2009/05. – Moscow: State University - Higher School of Economics. - 25 p.

In the work the fair division problem for two participants in presence of both divisible and indivisible items is considered. The set of all divisions is formally described; it is demonstrated that fair (in terms of Brams and Taylor) divisions, unlikely the case where all the items are divisible, not always exist. The necessary and sufficient conditions of existence of proportional and equitable division were found. Three interrelated modifications of the notion of fair division – profitably, uniformly and equitably fair divisions – were introduced. Computationally efficient algorithm for finding all of them was designed. The algorithm includes repetitive solutions of integer knapsack-type problems as its essential steps. The statements of the article are illustrated by various examples.

*Alexander Rubchinsky - State University – Higher School of Economics and International University "Dubna*

## 1. Introduction

In the middle nineties two American scientists – Steven Brams and Alan Taylor – suggested a fresh approach to widespread conflict situations. In these situations conflict consists of a family of separate disputable items (partial conflicts), and conflict resolution can be reduced to agreements about each of them. The main idea can be roughly presented as follows. Participants have *their own values* of *importance* of different items that form together a conflict. Because in most cases these values do not coincide completely, it is possible to achieve conflict resolution, such that both participants receive more than 50% of satisfaction *measured* in their own values.

### 1.1. Examples

The approach is carefully exposed in very comprehensive books [1, 2]. In order to clarify the approach two examples taken from [2] are considered here. Many other examples of real and hypothetical conflicts illustrating wide applicability of the approach can be found in the above mentioned books.

**Example 1. Divorse arrangement.** Ann and Ben are getting a divorce. The items that Bob and Carol had to divide were as follows:

A retirement account (pension), which, though substantial, will remain untouchable for several years; they are valuable for both but especially for Ann because Ben has more chance to make new account before his retirement.

A four-bedroom house, located close to Ben's job; therefore Ben values this house higher than Ann.

Country cottage that can be used at any season, preferable by Ann who intends to live there.

A portfolio of investments, which has lower monetary value than the pension but is all liquid assets.

Other, consisting of two cars and relatively expensive yacht highly valued by Ben.

In more detail the situation is described in [2]. Valuations of Ann and Ben of all the considered items are given in Table 1.

Table 1

| *Item* | *Ann* | *Ben* |
|---|---|---|
| Retirement account | 50 | 40 |
| House | 20 | 30 |
| Cottage | 15 | 10 |
| Portfolio | 10 | 10 |
| Other | 5 | 10 |
| Total | 100 | 100 |

Giving to everyone entire items, it is possible to suggest the following division:
For Ann: retirement account + other = 50 + 5 = 55;
For Ben: house + cottage + portfolio = 30 + 10 + 10 = 50.
Thus, satisfaction with this division is not less than half for both participants. The exact notion of optimal or fair division, suggested by Brams and Taylor, will be considered further.

**Example 2. Mergers.** Disagreements between businesses are common, especially when companies merge or are acquired. If each company cares more about different parts of an agreement, complex arrangements need to be worked out to satisfy both sides. One of the most elusive ingredients in the success of a merger is what deal makers euphemistically refer to as social issues – how power, position, and status will be allocated among the merging companies' executives. A failure to resolve these issues often leads to the destruction of shareholder wealth and the portrayal of top executives as petty corporate titans, unable to subordinate their selfish interests to the goal of promoting shareholder well-being.

Social issues concern the more ineffable matters of status, role, and prestige in the merged company, as opposed to "hard" financial factors. Even if a merger is ultimately consummated, as in the case of Boeing and McDonnel Douglas, a failure to agree on the resolution of social issues quickly wastes resources and the extremely valuable time of top corporate executives. The difficulty in forging cooperation between two management teams is perhaps inevitable, given the transformation that their relationship undergoes from the premerger to the post-merger period. After all, former adversaries, first in the marketplace and then at the negotiating table, are quite suddenly expected to work closely together and cooperate fully as their respective corporate entities attempt to meld into a single organization.

The following social issues are typical for companies' merger:
- the surviving company's name;
- the location of corporate headquarters;
- the split of the chairman and chief executive officer (CEO) positions;
- and, finally, which side will lay off some of its employees, particularly corporate executives, to eliminate overlapping operations or responsibilities (each company would prefer fewer of its own layoffs).

Suppose that the merging companies' executives negotiate over these issues in good faith. Thus, we are concerned with truly intractable issues that can be won or lost by either side without undermining the merger's objectives. Assume that each side distributes its 100 points across the issues, as follows:

Table 2

| Conflict items | Firm A | Firm B |
|---|---|---|
| 1. Name | 10 | 25 |
| 2. Headquarters | 20 | 35 |
| 3. President assignment | 15 | 20 |
| 4. CEO assignment | 25 | 10 |
| 5. Laying off | 30 | 10 |
| Total | 100 | 100 |

It is easy to see that in this hypothetical case each side can receive more than a half. For instance,
For firm A: item 4 + item 5 = 25 + 30 = 55;
For firm B: item 1 + item 2 + item 3 = 25 + 35 + 20 = 80.

However, this arrangement does not seem too fair: firm B receives (in its own values) significantly more than firm A receives (in its own values). A fair conflict resolution in this case is considered further in the article.

**1.2. Notion of fair division**

The considered in section 1.1 conflict situations, despite all the differences between them, clear demonstrate two key features, inherent to such kind of conflicts:
- a) conflict consists of several items (goods, issues, etc);
- b) participants have their own values of importance of every item.

These features form the basis of the above mentioned approach to conflict resolution. The task of conflict resolution is reduced to fair division – namely, to fair division of items importance among conflict participants correspondingly to their own evaluations. The fairness of division means the fulfillment of certain **formal conditions**:

1. *Proportionality*. Every one of participants thinks he or she received a portion that has a size or value of at least 50% in his or her own valuation.

2. *Equitability*. Each participant thinks that the portion he or she receives is worth the same, in terms of his or her valuation, as the portion that the other participant receives in terms of that participant's valuation.

3. *Efficiency*. A division is efficient (Pareto-optimal) if there is no other division that is strictly better for at least one participant and as good for another.

**1.3. Adjusted-winner procedure**

The authors of book [1] suggested surprisingly simple algorithm of fair division named the adjusted-winner (AW) procedure. The initial data consist of two sets of natural numbers: $a_1, \ldots, a_N$ and $b_1, \ldots, b_N$ that are valuations of participant A and B

$$\sum_{i=1}^{N} a_i = \sum_{i=1}^{N} b_i = 100.$$

Let us reorder all the items so that

$$a_1/b_1 \geq a_2/b_2 \geq \ldots \geq a_N/b_N. \quad (1)$$

The AW procedure of fair division can be presented as follows.

**Algorithm 1 (AW procedure).**

1. If $a_1 > \sum_{i=2}^{N} b_i$, then the 1st participant receives $x$-th share of item 1, or $xa_1$, while the 2nd participant receives $(1-x)$-th share of item 1 and all the other items 2, ..., N, or $(1-x)b_1 + \sum_{i=2}^{N} b_i$, where

$$x = 100/(a_1 + b_1);$$

the algorithm is completed.

2. If $\sum_{i=1}^{N-1} a_i \leq b_N$, then the 1st participant receives all the items 1, ..., N–1 and $x$-th share of item N, or $\sum_{i=1}^{N-1} a_i + xa_N$, while the 2nd participant receives $(1-x)$-th share of item N, or $(1-x)b_N$, where

$$x = 1 - 100/(a_N + b_N);$$

the algorithm is completed.

3. Starting with 1, increase $i$ up to some value $r$ satisfying conditions

$$\sum_{i=1}^{r-1} a_i \leq \sum_{i=r}^{N} b_i, \quad (2a)$$
$$\sum_{i=1}^{r} a_i > \sum_{i=r+1}^{N} b_i. \quad (2b)$$

4. If there is the equality in (2a), then the 1st participant receives items 1, 2, ..., r–1, or $\sum_{i=1}^{r-1} a_i$, while the 2nd participant receives items r, r+1, ..., N, or $\sum_{i=r}^{N} b_i$; the algorithm is completed.

5. If there is the inequality in (2a), then the 1st participant receives items 1, 2, ..., r–1 and $x$-th share of item $r$, or $\sum_{i=1}^{r-1} a_i + xa_r$, while the 2nd participant receives $(1-x)$-th share of item $r$ and items r+1, ..., N, or $(1-x)b_r + \sum_{i=r+1}^{N} b_i$, where

$$x = (\sum_{i=r}^{N} b_i - \sum_{i=1}^{r-1} a_i)/(a_r + b_r);$$

the algorithm is completed.

**Theorem 4.1** from book [1] states that a division constructed by this algorithm is a fair division, i.e. it possesses properties of proportionality, equitability and efficiency for arbitrary valuations $a_1, \ldots, a_N$ and $b_1, \ldots, b_N$.

**Example 1 (continuation).** Let us reorder items from Table 1 so that they satisfy condition (1):

Table 3

| Item | Ann | Ben |
|---|---|---|
| 1. Cottage | 15 | 10 |
| 2. Retirement account | 50 | 40 |
| 3. Portfolio | 10 | 10 |
| 4. House | 20 | 30 |
| 5. Other | 5 | 10 |
| Total | 100 | 100 |

Following algorithm 1, find a number $r$, satisfying (2). From Table 3 we have
$\sum_{i=1}^{1} a_i = 15$, $\sum_{i=2}^{5} b_i = 90$, $\sum_{i=1}^{2} a_i = 55$, $\sum_{i=3}^{5} b_i = 50$,
that means that $r = 2$ satisfies (2). In correspondence with step 5 of AW procedure, Ann receives cottage (15), and (5/6) of retirement account, while Ben receives portfolio (10), house (30), other (10),) and (1/6) of retirement account. Thus, everyone has 56.67% in terms of their own valuations, which is essentially more than a half (50%). Proportionality, equitability and efficiency of this division immediately follow from the above mentioned theorem. Note, that this fair division gives more for both participants than the division, considered at the beginning of this example.

**1.4. Notion of divisibility**

In [1], an item (good, issue) is called *divisible*, if it can be divided at any point along a continuum without destroying its value, and *indivisible*, if it cannot be divided without destroying its value. Of course, the notion of divisibility is an informal notion that must be discussed before division itself. Despite the significant (and sometimes crucial) importance of divisibility in many practical division problems, the analysis of this notion is beyond the scope of this work. However, one of the most attractive features of AW procedure consists in the following. This procedure guarantees that not more than one item must be divided; every other item is given to one of the participants entirely. This allows in many cases to avoid items division that is often difficult and informal task.

Because in AW procedure only one item (whose number is determined by algorithm 1) can be divided, this procedure gives a fair (i.e. proportional, equitable and efficient) division in all the cases, where this selected item is divisible **independently of divisibility of all the other items**.

Further we will consider the data about divisibility of items as initial data that cannot be discussed at the framework of this investigation. Of course, possibility (or impossibility) to divide some items can affect results of division, sometimes essentially. The next example gives a simple illustration of the notion of divisibility.

**Example 2 (continuation).** Let us reorder items from Table 2 so that they satisfy condition (1):

Table 4

| Conflict items | Firm A | Firm B |
|---|---|---|
| 1. Laying off | 30 | 10 |
| 2. CEO assignment | 25 | 10 |
| 3. President assignment | 15 | 20 |
| 4. Headquarters | 20 | 35 |
| 5. Name | 10 | 25 |
| Total | 100 | 100 |

Here, as everywhere in this material, values given by participants mean only relative importance of the corresponding items; they do not have any positive or negative sense.

It is natural to consider only the first item as a divisible one, while the other items cannot be divided (at least, arbitrary). Division possibilities in this case are carefully analyzed in [2]. Let us start with AW procedure without care about divisibility. The procedure gives items 1 and 2 to firm A, items 4 and 5 to firm B, and requires the division of item 3, i.e. president assignment, in proportion 5:2 between participants. Everyone receives 65,71% in terms of their own valuations.

In this case it is not difficult simply to consider all the possibilities and find a fair division satisfying the following condition: only the first item (laying off) can be divided. Firm A receives item 2 (25%), item 4 (20%), and (3∕4) of item 1 (22,5%), while firm B receives item 4 (35%), item 5 (25%) and (1∕4) of item 1 (2,5%). Both have 62,5%; it is naturally that taking into account indivisibility of some items can decrease the common gains; here we have 62,5% instead of 65,71%.

### 1.5. Informal statement of problem

**Example 3.** Let us consider a hypothetical division problem with two divisible and three indivisible items:

Table 5

| Items | A | B |
|---|---|---|
| 1 | 10 | 30 |
| 2 | 10 | 20 |
| 3 | 35 | 18 |
| 4 | 30 | 20 |
| 5 | 15 | 12 |

| Total | 100 | 100 |

Items 1 and 2 are divisible; items 3, 4 and 5 are indivisible.

Let us consider two different divisions. The 1st division: participant A receives indivisible item 3 (35%), indivisible item 5 (15%) and (2/3) of divisible item 2 (6,67%), while participant B receives indivisible item 4 (20%), divisible item 1 (30%) and (1/3) of divisible item 2 (6,67%); therefore each participant has 56,67%. The 2nd division: participant A receives indivisible item 3 (35%) and indivisible item 4 (30%), while participant B receives indivisible item 5 (12%), divisible item 1 (30%) and divisible item 2 (20%); therefore, A has 65% and B has 62%. This means that the 1st division is not a fair division because it is not efficient (65 >56,67 and 62 >56,67).

Example 3 will be formally analyzed further but it is clear that no fair division exists in the considered case. The first division looks fair, because it gives the same gain to both participants (and no other division gives greater equal gains). The second division looks fair, because it gives more to each participant (and cannot be improved for both of them: if one receives more, the other receives less).

The fact of the absence of fair (in the above mentioned sense) divisions in some situations, of course, is not a new one. However, the problem of fair division in cases where both types of items – divisible and indivisible – are present simultaneously up to now has not been formally stated and studied. In this connection the following problems arise:

- introduce reasonable formal modifications of the notion of fair division, adequate to both types of items;
- elaborate a computationally efficient (non-enumerative) method for finding division fair in the introduced senses;
- formulate and prove necessary and sufficient conditions of existence of proportional, equitable and fair divisions in terms of items evaluation by participants.

The article is devoted to solving these problems.

## 2. Basic notions, definitions, and statements

Let us start with more detailed notions. Assume there are $L$ divisible and $M$ indivisible items. Denote value of divisible and indivisible items for participant A by $a_1^d, \ldots, a_L^d, a_1^w, \ldots, a_M^w$ and for participant B by $b_1^d, \ldots, b_L^d, b_1^w, \ldots, b_M^w$. The sums of all the values are the same for each participant. It is not required that they are equal to 100. Instead of 100 it is possible to consider any natural number (denoted by $H$):

$$\sum_{i=1}^{L} a_i^d + \sum_{i=1}^{M} a_i^w = \sum_{i=1}^{L} b_i^d + \sum_{i=1}^{M} b_i^w = H. \qquad (3)$$

Cases $L = 0$ (no divisible items) and $M = 0$ (no indivisible items) are not excluded. The triple of numbers $S = \langle L, M, H \rangle$ will be called the *signature* of a division problem.

Thus, any division problem can be formally described by a triple $\langle a, b, S \rangle$ of three objects: $a = \langle a^d, a^w \rangle$, $b = \langle b^d, b^w \rangle$, $S = \langle L, M, H \rangle$, where

$$a^d = (a_1^d, \ldots, a_L^d), a^w = (a_1^w, \ldots, a_M^w), b^d = (b_1^d, \ldots, b_L^d), b^w = (b_1^w, \ldots, b_M^w),$$

S is the signature of the problem.

Furthermore, any division in a problem with signature S is presented as a couple $\langle x, \sigma \rangle$, where $x = (x_1, \ldots, x_L)$, $x_i$ are real numbers such that

$$0 \leq x_j \leq 1 \; (i = 1, \ldots, L),$$

and $\sigma = (\sigma_1, \ldots, \sigma_M)$,

$$\sigma_i \in \{0, 1\} \; (i = 1, \ldots, M).$$

In these terms participant A receives $(x_i)$-th part and participant B receives $(1-x_i)$-th part of divisible item $i$ ($i = 1, \ldots, L$); indivisible item $i$ is given to participant A, if $\sigma_i = 1$, and to participant B, if $\sigma_i = 0$ ($i = 1, \ldots, M$).

Let us consider an arbitrary division problem $\langle a, b, S \rangle$. Assume $\langle x, \sigma \rangle$ is a division in this problem. Assume

$$G_A^d(x) = \sum_{i=1}^{L} a_i^d x_i, \qquad (4a)$$

$$G_B^d(x) = \sum_{i=1}^{L} b_i^d (1 - x_i), \tag{4b}$$

$$G_A^w(\sigma) = \sum_{i=1}^{M} a_i^w \sigma_i, \tag{5a}$$

$$G_B^w(\sigma) = \sum_{i=1}^{M} b_i^w (1 - \sigma_i), \tag{5b}$$

$$G_A(x, \sigma) = G_A^d(x) + G_A^w(\sigma), \tag{6a}$$

$$G_B(x, \sigma) = G_B^d(x) + G_B^w(\sigma). \tag{6b}$$

Thus, $G(x, \sigma) = (G_A(x, \sigma), (G_B(x, \sigma))$ is the pair of gains received by participants as a result of division $\langle x, \sigma \rangle$. Pair $G^d(x) = (G_A^d(x), G_B^d(x))$ is the pair of gains of distribution $x$ of divisible items only, and pair $G^w(\sigma) = (G_A^w(\sigma), G_B^w(\sigma))$ is the pair of gains of distribution $\sigma$ of indivisible items only.

From the above notations it is clear that for any division $\langle x, \sigma \rangle$

$$G(x, \sigma) = G^d(x) + G^w(\sigma) \tag{7}$$

(the sum in (7) is the vector sum).

## 2.1. Properties of divisions

The properties of divisions mentioned in section 1.2 are expressed in these terms as follows.

***Efficiency*** of division $\langle x, \sigma \rangle$ means that for <u>any other division</u> $\langle y, \tau \rangle$ at least one of the following inequalities:

$$G_A(x, \sigma) > G_A(y, \tau),$$
$$G_B(x, \sigma) > G_B(y, \tau)$$

is true (none division gives more for one and at least the same for the other).

***Proportionality*** of a division $\langle x, \sigma \rangle$ means that

$$G_A(x, \sigma) \geq H/2,$$
$$G_B(x, \sigma) \geq H/2$$

(everyone receives at least half in his or her own values.

***Equitability*** of a division $\langle x, \sigma \rangle$ means that both participants receive the same (in their own values):

$$G_A(x, \sigma) = G_B(x, \sigma). \tag{8}$$

Remember that a division $\langle x, \sigma \rangle$ was named a ***fair division***, if it possesses properties of efficiency, proportionality and equitability. Here the simple interconnections between these properties are considered.

Assume

$U(S)$ be the set of all the division problems with a given signature S (the set of all the triples $\langle a, b, S \rangle$);
$E(S)$ be the set of all such problems that have efficient divisions,
$P(S)$ be the set of all such problems that have proportional divisions,
$Q(S)$ be the set of all such problems that have equitable divisions,
$F(S)$ be the set of all such problems that have fair divisions.
Let us start with the following simple

**Statement 1.** The connections between the considered sets is described by the following inclusions:

$$F(S) \subseteq Q(S) \subseteq P(S) \subseteq E(S) = U(S). \tag{9}$$

**Proof.** Let us go from right to left in (9).

1. Assume $\langle x, \sigma \rangle$ is an arbitrary division in an arbitrary problem $\langle a, b, S \rangle$ with a given signature S. Assume that in S $L > 0$ (at least one divisible item exists). If $\langle x, \sigma \rangle$ it is efficient itself that means that the considered problem belongs to $E(S)$ by definition. Otherwise, denote by $K(x, \sigma)$ the set of all the divisions that dominate $\langle x, \sigma \rangle$ or give the same gains. Because there is a finite set of different Boolean vectors $\tau$, the set $K(x, \sigma)$ is the union of a finite number of subsets $K_\tau(x, \sigma)$ consisting of divisions with the same indivisible part $\tau$. It is clear that all these sets $K_\tau(x, \sigma)$ are compact ones; hence, the corresponding sets of pairs $(G_A, G_B)$ are compact, too, (see formulae (4) – (6)). This implies that the set $R(x, \sigma)$ of all the gains, corresponding to division from $K(x, \sigma)$, is a compact one. Therefore there is at least one division $\langle y, \tau \rangle$, such that the corresponding pair $(G_A(y, \tau), G_B(y, \tau))$ maximizes $G_A + G_B$ over set $R(x, \sigma)$. Division $\langle y, \tau \rangle$ is an efficient division. If it is wrong that means that there is a division $\langle z, \upsilon \rangle$, dominating $\langle y, \tau \rangle$; hence, it dominates $\langle x, \sigma \rangle$. Then by the construction it must belong to $K(x, \sigma)$, but by the definition of $\langle y, \tau \rangle$ it is impossible. The contradiction proves the assertion in case $L > 0$. In case $L = 0$ all the considered sets are finite ones and the assertion is true.

2. The inclusion $P(S) \subseteq E(S)$ is proved analogously the previous one: any proportional division is efficient itself or it is dominated by an efficient division (this is proved exactly as in the previous part 1); the latest is proportional because every participant gains more or the same.

3. In order to prove the inclusion $Q(S) \subseteq P(S)$, let us introduce the useful notion of ***complement division***. For any division $\langle x, \sigma \rangle$ let us define the division $\langle \bar{x}, \bar{\sigma} \rangle$ (called the complement division) such that

$$\bar{x}_i = 1 - x_i \ (i = 1, \ldots, L),$$
$$\bar{\sigma}_i = 1 - \sigma_i \ (i = 1, \ldots, M).$$

Formulas (3) and (4) – (6) imply that

$$G_A(\bar{x}, \bar{\sigma}) = H - G_A(x, \sigma),$$
$$G_B(\bar{x}, \bar{\sigma}) = H - G_B(x, \sigma);$$

because $G_A(x, \sigma) = G_B(x, \sigma)$, at least one of two equitable divisions $\langle x, \sigma \rangle$ and $\langle \bar{x}, \bar{\sigma} \rangle$ is proportional that implies the required inclusion.

4. The inclusion $F(S) \subseteq Q(S)$ is true, because by definition of fair division it is equitable.

The statement 1 is proved.

Above mentioned theorem 4.1 from book [1] can be formulated as follows.

**Statement 2.** For division problems such that in their signatures $M = 0$ (i.e. any item is divisible) all the inclusions in (9) are equalities.

It is clear that even one indivisible item disturbs this beautiful picture. At the same time, fear divisions can exist even in the case $L = 0$ (i.e. any item is indivisible).

Let us introduce notions concerning an arbitrary division problem $\langle a, b, S \rangle$ itself (not its specific divisions). Denote

$$A_d(a, b, S) = \bigcup_x G^d(x), \tag{10}$$

where the union is taken over all the distributions $x$ of $L$ divisible items only;

$$A_w(a, b, S) = \bigcup_\sigma G^w(\sigma),$$

where the union is taken over all the distributions $\sigma$ of $M$ indivisible items only;

$$A(a, b, S) = \bigcup_{\langle x, \sigma \rangle} G(x, \sigma),$$

where the union is taken over all the divisions $\langle x, \sigma \rangle$ in problems with a given signature S.
The three defined sets – $A_d(a, b, S)$, $A_w(a, b, S)$, $A(a, b, S)$ – are sets of points in plane.
From the above notations and formula (7) it is clear that

$$A(a, b, S) = A_d(a, b, S) + A_w(a, b, S). \tag{11}$$

Finally, denote by $A_d^P(a, b, S)$, $A_w^P(a, b, S)$ and $A^P(a, b, S)$ Pareto-optimal parts of $A_d(a, b, S)$, $A_w(a, b, S)$ and $A(a, b, S)$, correspondingly.
Introduced notions immediately imply

**Statement 3.** Assume a division $\langle x, \sigma \rangle$ is efficient, i.e. $G(x, \sigma) \in A^P(a, b, S)$. Then $G^d(x) \in A_d^P(a, b, S)$ and $G^w(\sigma) \in A_w^P(a, b, S)$.

This statement means that for any efficient divisions $\langle x, \sigma \rangle$ its «projections» $x$ and $\sigma$ are efficient in the corresponding «projections» $A_d(a, b, S)$ and $A_w(a, b, S)$ of set $A(a, b, S)$. Set $A(a, b, S)$ can be named «attainability set» of the considered division problem $\langle a, b, S \rangle$.

In order to illustrate the introduced notions, consider the following example (see section 1.5).

**Example 3. Continuation.** 1. The initial data for this case is given in Table 5. The two first rows in this table describe divisible items that are rewritten in Table 6:

Table 6

| Items | A | B |
|-------|-----|-----|
| 1 | 10 | 30 |
| 2 | 10 | 20 |
| Total | 20 | 50 |

Set $A_d$ ($a$, $b$, S) of all the points, corresponding to all the possible distributions of these divisible items, is the set of all points presented as

$$(10x+10y, 30(1-x)+20(1-y)), \qquad (12)$$

where

$$0 \leq x \leq 1, 0 \leq y \leq 1. \qquad (13)$$

In (12), (13) $x$ and $y$ are the parts of divisible items 1 and 2, received by participant A.

Moving pair ($x$, $y$) along the way $(0,0) \to (0,1) \to (1,1) \to (1,0) \to (0,0)$ and calculating corresponding points by formula (12), we find the quadrilateral shown in Fig.1. It coincides with $A_d$ ($a$, $b$, S). Its Pareto-optimal part $A_d^P(a, b, S)$ is shown in the figure.

2. The three last rows in this table 5 describe indivisible items that are rewritten in Table 7. There are 8 possibilities of distributions of three indivisible items between two participants that can be presented by 8 Boolean vectors $\sigma = (\sigma_1, \sigma_2, \sigma_3)$. Using formulae (5) we find that set $A_w(a, b, S)$ in the considered case consists of 8 points

$$(35\sigma_1 + 30\sigma_2 + 15\sigma_3, 18(1-\sigma_1) + 20(1-\sigma_2) + 12(1-\sigma_3)),$$

where ($\sigma_1$, $\sigma_2$, $\sigma_3$) are all the Boolean vectors with 3 components. Substituting ($\sigma_1$, $\sigma_2$, $\sigma_3$) consecutively with their values from (0,0,0) till (1,1,1) easily find 8 points in the plane:

$$A_w(a, b, S) = \{(0, 50), (15, 38), (30, 30), (45, 18), (35, 32), (50, 20), (65, 12), (80, 0)\}. \qquad (14)$$

Among 8 points (14) there are two dominated: (30, 30) is dominated by (35, 32), and (45, 18) is dominated by (50, 20). Therefore, $A_w^P(a, b, S)$ consists of 6 points:

$$A_w^P(a, b, S) = \{(0, 50), (15, 38), (35, 32), (50, 20), (65, 12), (80, 0)\}.$$

3. The attainability set $A(a, b, S)$ in the considered example is easily constructed using (11) and explicit presentation of sets $A_d$ ($a$, $b$, S) and $A_w$ ($a$, $b$, S), given in Fig.1 and 2. This set $A(a, b, S)$ is marked grey in Fig.3. The Pareto-optimal part $A^P(a, b, S)$ of the attainability set $A(a, b, S)$ is shown in Fig.4.

By virtue of statement 3, it is enough to add to each point belonging to the finite set $A_w^P(a, b, S)$, shown in Fig.2, the set $A_d^P(a, b, S)$, shown in Fig.1 (adding here means coordinate adding of two-dimension vectors).

In Fig.4 the different parts of Pareto-optimal border $A^P(a, b, S)$ of the attainability set $A(a, b, S)$ are marked bold; a white circle at the end of a part means that the corresponding point belongs to the part; the absence of a circle means that the corresponding end does not belong to $A^P(a, b, S)$. The point marked by a black square presents an equitable division with maximal possible gains 56,67. This division is not a fair one, because it is not efficient: the point (65, 62) dominates the marked equitable division as well as any other equitable division (65>56,67 and 62>56,67). Therefore, in the considered problem there are no fair divisions at all. Note that the properties of divisions are determined by geometry of Pareto set $A^P(a, b, S)$.

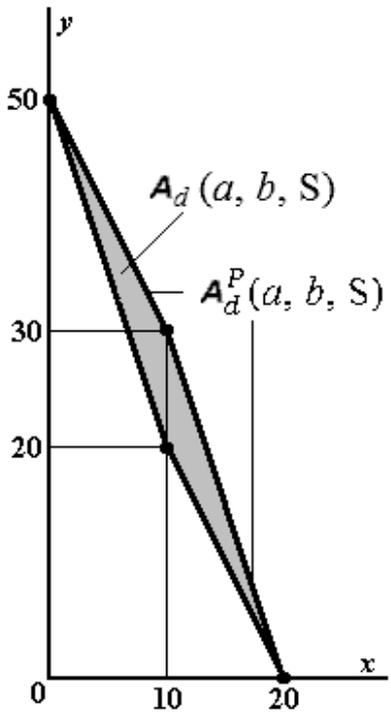

| Table 7 | | |
|---|---|---|
| Items | A | B |
| 3 | 35 | 18 |
| 4 | 30 | 20 |
| 5 | 15 | 12 |
| Total | 80 | 50 |

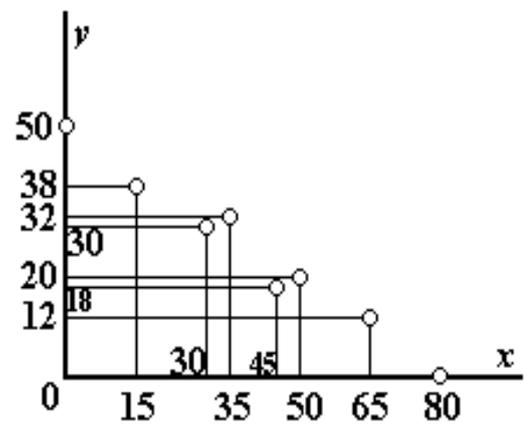

Fig.1

Fig.2

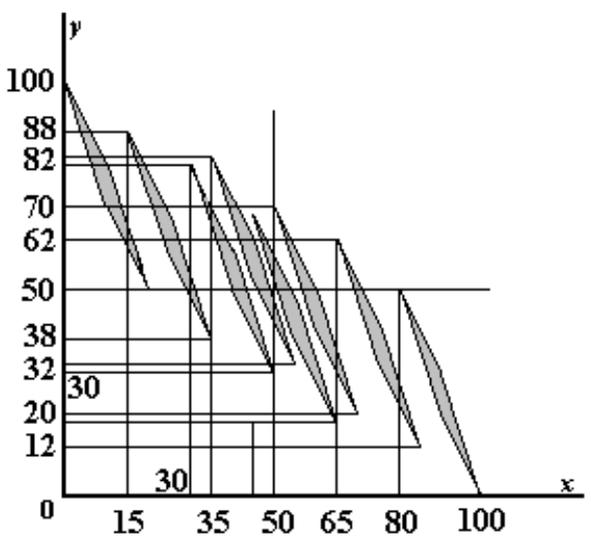

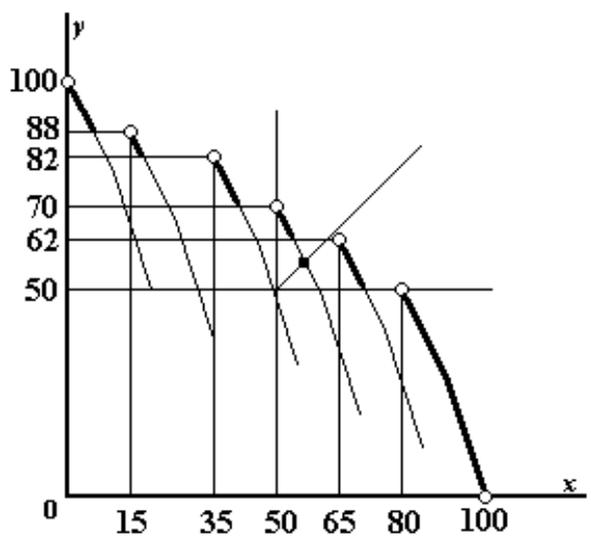

Fig.3

Fig.4

## 2.2. General structure of $A_d(a, b, S)$

In Example 3 (see Fig.1) set $A_d(a, b, S)$ is a center-symmetrical convex polygon, whose all the edges have negative slope. The same is true for arbitrary division problem $\langle a, b, S \rangle$. Let us consider the structure of $A_d(a, b, S)$ in general case in more detail.

Assume

$$s_1 = \sum_{i=1}^{L} a_i, \qquad (15a)$$

$$s_2 = \sum_{i=1}^{L} b_i \qquad (15b)$$

(index $d$ is omitted for simplicity). To construct the set, assume that values are ordered as in (1):

$$a_1/b_1 \geq a_2/b_2 \geq \ldots \geq a_L/b_L. \qquad (16)$$

It is clear that set $A_d(a, b, S)$ is a convex polygon.

Let us consider the unit cube in $L$-dimension space. Any point $x = (x_1, ..., x_L)$ belonging to this cube determines point $f(x)$ belonging to polygon $A_d(a, b, S)$:

$$f(x) = (\sum_{i=1}^{L} a_i x_i, \sum_{i=1}^{L} b_i(1 - x_i))$$

(see formulae (4)) and any point of the polygon coincides with $f(x)$ for some $x$. Let us consider all the vertices of the unit cube starting with $(0,0,...,0)$ till $(1,1,...,1)$. The images of these vertices cover completely all the vertices of polygon $A_d(a, b, S)$. The vertex $(0,0,...,0)$ correspond to the left point $(0, s_2)$. Consider the edges connecting vertex $(0,0,...,0)$ to the $L$ adjacent vertices of unit cub: $(1,0,...,0)$, $(0,1,...,0)$, ..., $(0,0,...,1)$. Corresponding segments on the plane have the common left end $(0, s_2)$ and slopes $b_i/a_i$ ($i = 1, ..., L$). Order (16) means that the minimal possible slope has the segment connecting point $(0, s_2)$ to the point $(a_1, s_2-b_1)$, because its inverse value $a_1/b_1$ is the maximal. That means that this segment is a part of border of $A_d(a, b, S)$. Considering the next vertex $(a_1, s_2-b_1)$, analogously find the next border point $(a_1 + a_2, s_2-b_1-b_2)$, and so on, up to the last point $(s_1, 0)$ (see (15)). Thus, set $A_d^P(a, b, S)$ is the broken line whose $k$-th point is:

$(\sum_{i=1}^{k} a_i, \sum_{i=k+1}^{L} b_i)$ ($k = 1, ..., L$);

The 0-th point is $(0, s_1)$. The geometrical illustration is given in Fig.5. The same reasoning implies that the low border of the polygon consists of the same segments in reverse order. Therefore, point $(s_1/2, s_2/2)$ is the center of symmetry of the polygon.

In the case $M = 0$ there are only $L$ divisible items, and the reasoning of this section gives another proof of statement 2 (theorem 4.1 from [1]). Indeed, in this case $s_1 = s_2 = H$, and intersection of line $x = y$ with broken line $A_d^P(a, b, S)$ determines the same fair division.

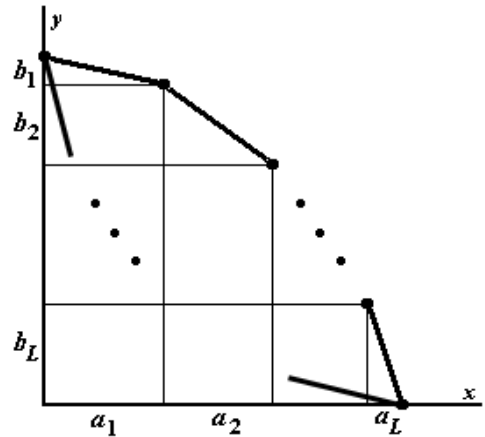

Fig.5

The established structure of set $A_d(a, b, S)$ implies several important corollaries (formulated further as statements):

**Statement 4.** Assume $\langle x, \sigma \rangle$ is an arbitrary efficient division. Then not more than one item must be divided; every other item is given to one of the participants entirely.

**Proof.** We have (see (7)) $G(x, \sigma) = G^d(x) + G^w(\sigma)$ and by virtue of statement 3 $G^d(x) \in A_d^P(a, b, S)$. That means that point $G(x, \sigma)$ belongs to broken line $G^w(\sigma) + A_d^P(a, b, S)$. Every vertex on this line presents distribution of all the divisible items as whole (see the previous construction), and any point inside a segment between two vertices presents division of the item, corresponding to this segment.

**Statement 5.** Assume $\langle x, \sigma \rangle$ is a maximal equitable division (i.e. an equitable division with maximal possible common gain). Then not more than one item must be divided; every other item is given to one of the participants entirely.

**Proof.** Because $\langle x, \sigma \rangle$ is an equitable division, point $G(x, \sigma)$ must belong to line $x = y$. Consider intersection of line $x = y$ with any polygon $G^w(\sigma) + A_d(a, b, S)$. If point $G(x, \sigma)$ does not belong to $G^w(\sigma) + A_d^P(a, b, S)$, then both coordinates of $G(x, \sigma)$ can be increased, i.e. $\langle x, \sigma \rangle$ is not a maximal equitable division. Therefore, it belongs to broken line $G^w(\sigma) + A_d^P(a, b, S)$ that implies the assertion by the same reasoning as at the end of the previous prove.

## 3. Modifications of notion of fair division

Now we introduce the essential notions of this work – modifications of notion of fair division. In the division problem from Example 3 it is impossible to provide efficiency and equitability simultaneously – we can receive the equitable division that is not efficient or the efficient division that is not equitable (see points in Fig.4). It is not a pathologic exception: it really happens in several examples given further in section 6.

These examples led us to the following formal definitions.

A division $\langle x, \sigma \rangle$ in problem $\langle a, b, S \rangle$ is ***profitably fairy*** if it is
- proportional;
- efficient;
- maximizes expression $\min\{G_A(x, \sigma), G_B(x, \sigma)\}$ over the set of all the divisions $\langle x, \sigma \rangle$.

A division $\langle x, \sigma \rangle$ in problem $\langle a, b, S \rangle$ is ***uniformly fairy*** if it is
- proportional;
- efficient;
- minimizes expression $|G_A(x, \sigma) – G_B(x, \sigma)|$ over the set of all such divisions $\langle x, \sigma \rangle$.

A division $\langle x, \sigma \rangle$ in problem $\langle a, b, S \rangle$ is ***equitably fairy*** if it is
- proportional;
- equitable;
- maximizes common gain $G$ over all such divisions $\langle x, \sigma \rangle$.

Thus, profitably fair divisions maximize minimal profits of participants but do not care about equality; uniformly fair divisions try to make profits as equal (uniform) as possible, yet under efficiency restriction; finally, equitably fairy divisions provide equality of profits but do not care about efficiency. There is an evident connection between these different kinds of fairness; let us formulate it as an assertion.

**Statement 6.** Any division is fair if and only if it is profitably, uniformly, and equitably fair.

Let us consider a polygon, denoted by $A(\sigma)$:
$$A(\sigma) = G^w(\sigma) + A_d,$$
where $\sigma$ is an arbitrary distribution of indivisible items (see (5)), polygon $A_d$ is described in section 2.2. Remember that the union of all these polygons (over all Boolean vectors $\sigma$) forms the attainability set $A(a, b, S)$.

**Statement 7.** If polygon $A(\sigma)$ is completely below (above) line $x = y$, then its left (right) vertex maximizes expression
$$\min\{G_A(x, \sigma), G_B(x, \sigma)\}$$
and minimizes expression
$$|G_A(x, \sigma) – G_B(x, \sigma)|$$
over the polygon.

**Proof.** It is enough to consider any line with a negative slope, because slopes of all the edges of the polygon are negative. For the second expression the geometric illustration of this evident fact is given in Fig.6. For the first expression it is evident because minimal $y$-coordinate of B is less than of A.

**Statement 8.** If line $x = y$ intersects polygon $A(\sigma)$, then in intersection of this line and Pareto border of the polygon $A(\sigma)$ the same expressions reach maximal and minimal values over the polygon, so that the value of the second expression is equal to 0, and $G_A(x, \sigma) = G_B(x, \sigma)$}

This statement immediately follows from the previous definitions and constructions.

Modified definitions of fair divisions together with Statements 4 and 5 imply

**Statement 9.** Any profitably, any uniformly, and any equitably fair division has the property: not more than one item must be divided; every other item is given to one of the participants entirely.

It is clair that fair divisions also have the same property.

### 4. Algorithms
Statements 6 – 8 allow elaborating of computationally efficient algorithm of finding profitably, uniformly, and equitably fair divisions. All of them are found by this algorithm. Before describe the algorithm, let us consider the following useful procedure.

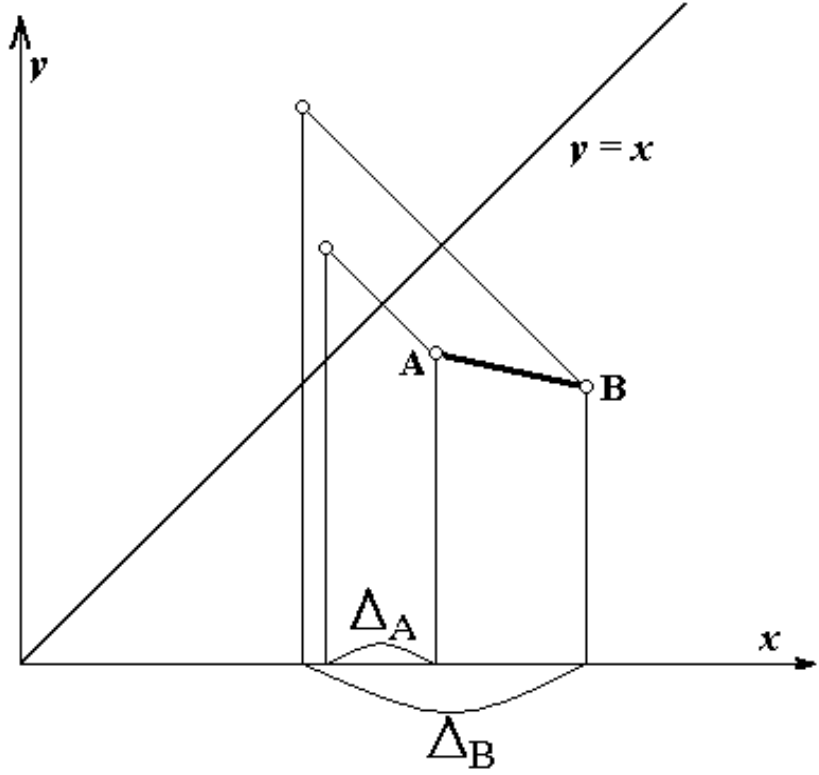
Fig.6

## 4.1. Constructing $A_w^P(a, b, S)$

Definition (5) means that set $A_w(a, b, S)$ consists of all points with coordinates

$$( \sum_{i=1}^{M} a_i \sigma_i, \ \sum_{i=1}^{M} b_i(1 - \sigma_i) ),$$

where $\sigma = (\sigma_1, \ldots, \sigma_M)$ is an arbitrary Boolean vector with $M$ components (index $w$ is omitted for simplicity).

Assume $A_x$ be the projection of set $A_w(a, b, S)$ to $x$-axis. In Example 6 $A_x$ consists of 8 numbers: $A_x =$ {0, 15, 30, 35, 45, 50, 65, 80} (see (14)). By definition of projection for any $x \in A_x$ there is at least one number $y$ such that

$$(x, y) \in A_w(a, b, S). \tag{17}$$

Denote the maximal $y$ satisfying (17) by $y(x)$ and define set $A$ as follows:
$$A = \{(x, y(x)) \mid x \in A_x \}.$$

By the definition of $A$ it is clear that

$$A_w^P(a, b, S) \subseteq A \subseteq A_w(a, b, S)$$

and

$$A_w^P(a, b, S) = A^P;$$

it means that the considered problem is reduced to construction of $A^P$. Note, that in Example 3 the second inclusion is equality.

In order to select the Pareto set $A^P$ from $A$ it is enough simply to start with maximal number from $A_x$ and go left, eliminating all the pairs in which $y$-components does not exceed the previous one. Therefore, the considered problem is reduced to the <u>problem of finding $A$</u>.

Assume

$$s_1 = \sum_{i=1}^{M} a_i,$$
$$s_2 = \sum_{i=1}^{M} b_i$$

and consider the following family of optimization problem relatively Boolean variables $\sigma_1, \ldots, \sigma_M$:

$$\sum_{i=1}^{M} b_i(1 - \sigma_i) \to \max \tag{18}$$

subject to

$$\sum_{i=1}^{M} a_i \sigma_i = k \quad (k = s_1,\ s_1-1,\ s_1-2,\ \ldots,\ 1,\ 0). \tag{19}$$

By the construction, for all $k$ that do not belong to $A_x$, equation (19) is incompatible. For $k = 0$ equation (19) is compatible: the corresponding value of goal function (18) is $s_2$.

Assume

$$\tau_i = 1 - \sigma_i \ (i = 1, 2, \ldots, M)$$

and substitute in (18), (19) $\sigma_i$ with $1 - \tau_i$ ($i = 1, 2, \ldots, M$). We receive the equivalent family of optimization problem relatively Boolean variables $\tau_1, \ldots, \tau_M$:

$$\sum_{i=1}^{M} b_i \tau_i \to \max \tag{20}$$

subject to

$$\sum_{i=1}^{M} a_i \tau_i = k \quad (k = 0, 1, \ldots, s_1). \tag{21}$$

Thus, the considered in this section problem of $A_w^P(a, b, S)$ construction is reduced to the solution of the family of optimization problems (20), (21). If for some $k$ a solution exists, it gives a point of $A$ corresponding to the following items distribution: participant A (B) receives item $i$, if $\tau_i = 0$ ($\tau_i = 1$).

Consider now family (20), (21) of optimization problems. The problems differ from well-known knapsack problem only in the following: instead of weight restrictions ($\leq$) there are equalities (=) in conditions (21). However, they can be solved (analogously knapsack problem) by dynamic programming. Let us consider the situation in more detail.

Denote the optimization problem

$$\sum_{i=1}^{p} b_i \tau_i \to \max$$

subject to

$$\sum_{i=1}^{p} a_i \tau_i = k$$

by $Z(k, p)$, and the optimal value of its goal function by $F(k, p)$. Then Bellman equation for the problem is:

$$F(k, p) = \max \{F(k, p-1), F(k-a_p, p-1) + b_p\} \ (k = 0, 1, \ldots, s_1;\ p = 2, \ldots, M);$$

if the maximal value here is $F(k, p-1)$, that means that in the optimal solution $\tau_p = 0$; otherwise, in the optimal solution $\tau_p = 1$.

Finally, initial value are determined by formula

$$F(k, 1) = \begin{cases} b_1, & \text{if } a_1 = k \\ -\infty, & \text{otherwise} \end{cases} \quad k = 0, 1, \ldots, s_1.$$

The solution of optimization problem $Z(k, p)$ for $p = M$ corresponds to the solution of the initial problem (20), (21). If the goal function for some $k$ is equal to $-\infty$, then for this $k$ equation (21) is incompatible.

Thus, in this section the efficient and relatively simple algorithm of $A_w^P(a, b, S)$ construction is described. In more detail: by the construction, the algorithm is efficient relatively number $H$ – the third component of the signature of the problem. Note, that the number of undominated points does not exceed $H$.

## 4.2. Checking dominance

In order to find profitably, uniformly, and equitably fair divisions it is necessary to know about arbitrary point $(x, y)$ belonging to attainability set $A(a, b, S)$, whether it corresponds to an efficient division, i.e. whether it belongs to Pareto-optimal part $A^P(a, b, S)$ of attainability set $A(a, b, S)$. Statement 3 immediately implies that $A^P(a, b, S)$ is a subset of the following set:

$$A^P(a, b, S) \subseteq \bigcup_{i=1}^{w} Z_i, \tag{22}$$

where

$$Z_i = W_i + A_d^P(a, b, S) \ (i = 1, \ldots, w). \tag{23}$$

In (23) points $W_i$ ($i = 1, \ldots, w$) form set $A_w^P(a, b, S)$ of different undominated pairs of gains from distribution of indivisible items only; they are found by the algorithm from section 4.1.

Remember that $A_d^P(a, b, S)$ is a broken line with $L$ edges, which is an upper part of border of a convex polygon $A_d(a, b, S)$ (see section 2.2 and Fig.5). By the construction of $Z_i$ all their graphs are received as shifts of the same graph $A_d^P(a, b, S)$ by vectors $W_i$.

Inclusion (22) implies that a point $(x, y)$ is undominated into $A(a, b, S)$ if and only if it does not dominated by any point belonging to $\bigcup_{i=1}^{w} Z_i$. Therefore, it is enough to check, whether point $(x, y)$ is dominated by a point belonging to one of known broken lines $W_i + A_d^P(a, b, S)$ ($i = 1, \ldots, w$). Hence, the problem is reduced to checking dominance for every broken line (23) separately. The last problem finally is reduced to the analogous problem for every of $L$ edges forming one line; and the solution of this problem is evident.

It is clear that the bulk of all these operations is computationally efficient (relatively parameters $w$ and $L$).

### 4.3. Constructing equitable fair division

In this section an algorithm for construction of an equitable fair division is suggested. Division $\langle x, \sigma \rangle$ is an equitable fair division if it is a solution of optimization problem

$$\sum_{i=1}^{L} a_i^d x_i + \sum_{i=1}^{M} a_i^w \sigma_i \to \max \tag{24}$$

subject to

$$\sum_{i=1}^{L} a_i^d x_i + \sum_{i=1}^{M} a_i^w \sigma_i = \sum_{i=1}^{L} b_i^d (1 - x_i) + \sum_{i=1}^{M} b_i^w (1 - \sigma_i). \tag{25}$$

Equality (25) for gains follows immediately from (8), taking into account (4) – (6).

Statement 9 allows reducing the considered problem to the analogous problem with exactly one indivisible item. In more detail: let us consider the problem with the same values of participants and additional condition: only item $k$ (among divisible in the initial problem items 1, 2, ..., $L$) is divisible; all the other items are indivisible. That means that in equality (25) variable $x_i \in \{0,1\}$ ($i=1, \ldots, M$; $i \neq k$). Denote this problem by $Z(k)$. Using introduced in section 2 notion of signature, the signature of $Z(k)$ is $\langle 1, L+M-1, H \rangle$ (instead of initial signature $S = \langle L, M, H \rangle$).

Denote the value of goal function (24) in problem $Z(k)$ by $F(k)$. Assume

$$F(k^*) = \max_{1 \leq k \leq M} F(k);$$

then the corresponding to this number $k^*$ division $\langle x, \sigma \rangle$ is an equitable fair division in the considered initial division problem $\langle a, b, S \rangle$.

Thus, the problem of construction of an equitable fair division is reduced to the same problem in situations where only one item is divisible.

**4.3.1. Case of one divisible item.** In order to simplify notations, in this section a division problem is described by a couple $\langle a, b \rangle$, where $a = (a_0, a_1, \ldots, a_N)$, $b = (b_0, b_1, \ldots, b_N)$ are values of participants; item 0 is only divisible item; all the other items are indivisible. Any division is presented by a couple $\langle x, \sigma \rangle$, where

$$0 \leq x \leq 1; \tag{26}$$

$\sigma = (\sigma_1, \ldots, \sigma_N)$ is a Boolean vector. In these notations optimization problem (24), (25) is written as

$$x a_0 + \sum_{i=1}^{N} a_i \sigma_i \to \max \tag{27}$$

subject to

$$x a_0 + \sum_{i=1}^{N} a_i \sigma_i = (1-x) b_0 + \sum_{i=1}^{N} b_i (1 - \sigma_i). \tag{28}$$

Condition (28) can be rewritten equivalently as

$$\sum_{i=1}^{N} a_i \sigma_i - \sum_{i=1}^{N} b_i (1 - \sigma_i) = (1-x) b_0 - x a_0. \tag{29}$$

Let us consider the double inequality

$$-a_0 \leq \sum_{i=1}^{N} a_i \sigma_i - \sum_{i=1}^{N} b_i (1 - \sigma_i) \leq b_0. \tag{30}$$

**Statement 10.** Couple $\langle x, \sigma \rangle$ satisfies conditions (26), (28) if and only if Boolean vector $\sigma$ satisfies double inequality (30).

**Proof.** 1. For $x = 0$ $(1-x)b_0 - xa_0 = b_0$, for $x = 1$ $(1-x)b_0 - xa_0 = -a_0$, which implies that for any $x \in [0,1]$ right-hand side of (29) belongs to segment $[-a_0, b_0]$. Therefore, for any couple $\langle x, \sigma \rangle$ satisfying (26), (28) vector $\sigma$ satisfies double inequality (30).

2. Assume Boolean vector $\sigma$ satisfies double inequality (30). Let

$$z = \sum_{i=1}^{N} a_i \sigma_i - \sum_{i=1}^{N} b_i (1 - \sigma_i). \tag{31}$$

By virtue of (30)

$$-a_0 \leq z \leq b_0. \tag{32}$$

Define a number $x$ from the equality

$$z = (1-x)b_0 - xa_0, \tag{33}$$

implying

$$x = \frac{b_0 - z}{b_0 + a_0}; \tag{34}$$

from (34) and right inequality in (32) we have $0 \leq x$, and from (34) and left inequality in (32) we have $x \leq 1$; together they coincide with (26).

Equality (28) follows from substitution in (31) $z$ with right-hand side of (33) and rearrangement of the expression. Statement is proved.

Assume

$$c_i = a_i + b_i \, (i = 1, \ldots, N), \tag{35}$$

**Statement 11.** Division problem $\langle a, b \rangle$ (with one divisible item) has an equitable division if and only if the following systems of two linear inequalities relatively Boolean variables $\sigma_1, \ldots, \sigma_N$

$$H - (a_0 + b_0) \leq \sum_{i=1}^{N} c_i \sigma_i \leq H. \tag{36}$$

is compatible.

**Proof.** By virtue of statement 10 equitability of a division is equivalent to existence of solutions of double inequality (30). Rewrite (30) as

$$-a_0 \leq \sum_{i=1}^{N}(a_i + b_i)\sigma_i - \sum_{i=1}^{N} b_i \leq b_0.$$

Adding to every part $\sum_{i=1}^{N} b_i$, we receive

$$-a_0 + \sum_{i=1}^{N} b_i \leq \sum_{i=1}^{N}(a_i + b_i)\sigma_i \leq b_0 + \sum_{i=1}^{N} b_i. \tag{37}$$

Right-hand side of (37) is the sum of all the values of participant B, and, hence, is equal to $H$. Left-hand side of (36) is equal to $-a_0 + (H - b_0) = H - (a_0 + b_0)$. Taking into account (35), finally write (37) as (36).

By the reverse way (36) can be rewritten as (30). By virtue of statement 10 double inequality (30) is equivalent to equitability of a division. The statement is proved.

**Statement 12.** An equitable division $\langle x, \sigma \rangle$ is a solution of optimization problem (27) subjected to (26), (28) if and only if Boolean vector $\sigma$ is a solution of optimization problem

$$\sum_{i=1}^{N} \sigma_i \frac{b_0 a_i - a_0 b_i}{b_0 + a_0} + \frac{H a_0}{b_0 + a_0} \to \max \tag{38}$$

subject to (36).

**Proof.** It is enough to express $x$ through $\sigma$ from (28), substitute $x$ with this expression in goal function (27) and rearrange the formula.

Now we can multiply goal function (38) by $(b_0 + a_0)$, subtract $Ha_0$, and finally receive

**Statement 13.** The considered problem of construction of equitable division in a problem with one divisible item is reduced to optimization problem relatively Boolean variables $\sigma_1, \ldots, \sigma_N$

$$\sum_{i=1}^{N} d_i \sigma_i \to \max \tag{39}$$

subject to (36), where $d_i = b_0 a_i - a_0 b_i$ $(i = 1, 2, \ldots, N)$ are integer numbers.

Problem (39), (36) is replaced by the family of problems with the same goal function (39) and condition

$$\sum_{i=1}^{N} c_i \sigma_i = k \quad (k = H-(a_0 + b_0), H-(a_0 + b_0)+1, \ldots, H); \tag{40}$$

they are solved as problems (20), (21) in section 4.1 (by dynamic programming).

Thus, the reasoning in this section shows that an equitable fair division in general case can be found by the suggested computationally efficient method. Pay attention that solving problem (39), (40) we automatically check compatibility of condition (40) for every $k = H-(a_0 + b_0)$, $H-(a_0 + b_0)+1, \ldots, H$; the incompatibility for all $k$ means that the considered division problem does not have equitably fair divisions.

### 4.4. Essential algorithm

The essential algorithm of finding profitably, uniformly, and equitably fair divisions is based on Statements 7 and 8 and the algorithms from section 4.1, 4.2 and 4.3.

The main steps of the essential algorithm briefly are described as follows.

<u>Step 1</u>. Assume sg = 1. Construct an equitably fair division $\langle x, \sigma \rangle$ by the algorithm from section 4.3. If no equitably fair division exists, assume sg = 0 and go to Step 3. Otherwise, assume $e = G_A(x, \sigma)$ and go to Step 2.

<u>Step 2</u>. Check dominancy of point $(e, e)$ (where $e$ is found in step 1) by the algorithm, described in section 4.2. If this point $(e, e)$ is undominated, then division $\langle x, \sigma \rangle$ is simultaneously profitably, uniformly and equitably fair division. The algorithm stops.

<u>Step 3</u>. Construct set $A_w^P(a, b, S) = \{W_1, W_2, \ldots, W_w\}$ by algorithm from section 4.1.

<u>Step 4</u>. Construct set (23) of broken lines $Z_i = W_i + A_d^P(a, b, S)$ $(i = 1, \ldots, w)$.

<u>Step 5</u>. Assume P = ∅, U = ∅ (initial assignments for current sets of profitably and uniformly fair divisions).

<u>Step 6</u>. For $i = 1, \ldots, w$ do the following:

<u>Step 6.1</u>. If $Z_i$ is completely below line $x = y$, denote the division corresponding to its left vertex by $\langle z, \tau \rangle$ and go to Step 6.3. Otherwise, go to Step 6.2.

<u>Step 6.2</u>. If $Z_i$ is completely above line $x = y$, denote the division corresponding to its right vertex by $\langle z, \tau \rangle$ and go to Step 6.3. Otherwise, continue the cycle in Step 6.

<u>Step 6.3</u>. Assume P = P ∪{$\langle z, \tau \rangle$}, U = U ∪{$\langle z, \tau \rangle$}; continue the cycle in Step 6.

<u>Step 7</u>. Assume Q = P ∪ U.

<u>Step 8</u>. Eliminate from Q all the divisions $\langle z, \tau \rangle$, such that $G_A(z, \tau)$ or $G_B(z, \tau)$ is less then $H/2$. If Q is empty, then profitably fair divisions, uniformly fair divisions, and equitably fair divisions in the considered problem do not exist. The algorithm stops. Otherwise, go to Step 9.

(Comment: if Q is empty, let us consider two cases: sg = 0 and sg = 1 (see Step 1). In the case sg = 0 no equitably fair division exists; in the case sg = 1 the equitably fair division $\langle x, \sigma \rangle$, found in Step 1, is the efficient division, but in this case the algorithm had to stop in Step 2).

<u>Step 9.</u> Denote by X the set of all the points $(G_A(z, \tau), G_B(z, \tau))$, where $\langle z, \tau \rangle \in Q$; denote by $X^P$ Pareto-optimal part of X and by $Q^P$ the set all the divisions, whose gains $(G_A(z, \tau), G_B(z, \tau))$ belong to $X^P$; denote by $q$ the cardinality of $X^P$.

<u>Step 10</u>. For $i = 1, \ldots, q$ do the following:

<u>Step 10.1</u>. Checking dominance of $i$-th point $(x_i, y_i)$ from $X^P$ by algorithm from section 4.2.

<u>Step 10.2</u>. If $(x_i, y_i)$ is dominated, then assume $X^P = X^P \setminus \{(x_i, y_i)\}$; continue.

<u>Step 11.</u> Maximize the expression min$\{x, y\}$ over set of points $X^P$. The corresponding division is a profitably fair one.

<u>Step 12.</u> Minimizes the expression $|x - y|$ over set of points $X^P$. The corresponding division is a uniformly fair one.

<u>Step 13.</u> The algorithm stops, because profitably, uniformly and equitably fair divisions are found (correspondingly, in Steps 11, 12 and 1).

## 5. Existence conditions in division problems

In this section the last of the mentioned at the end of section 1.5 problems is considered. Namely, though the diagnostic is included into essential algorithm described in section 4.4, the existence or absence of some properties can be established easier, without full completion of all the operations. We establish necessary and sufficient conditions of existence of proportional and equitable fair divisions in terms of items evaluation by participants.

Consider an arbitrary problem $\langle a, b, S\rangle$, where $S = \langle L, M, H\rangle$. That means that there are $L$ divisible items, $M$ indivisible items and sums of participants' values are equal to $H$. Assume

$$S_1 = \sum_{i=1}^{L} a_i^d,\ S_2 = \sum_{i=1}^{L} b_i^d,\ T_1 = \sum_{i=1}^{L} a_i^w,\ T_2 = \sum_{i=1}^{L} b_i^w. \tag{41}$$

Let us associate with any problem $\langle a, b, S\rangle$ the following systems of linear inequality relatively Boolean variables $\sigma_1, \ldots, \sigma_M$:

$$\begin{cases} \sum_{i=1}^{M} \sigma_i a_i^w \leq H/2, \\ \sum_{i=1}^{M} \sigma_i b_i^w \geq H/2; \end{cases} \tag{42}$$

$$\begin{cases} \sum_{i=1}^{M} \sigma_i a_i^w \geq H/2, \\ \sum_{i=1}^{M} \sigma_i b_i^w \leq H/2; \end{cases} \tag{43}$$

$$\begin{cases} \sum_{i=1}^{M} c_i \sigma_i \leq H, \\ \sum_{i=1}^{M} c_i \sigma_i \geq H - (S_1 + S_2), \end{cases} \tag{44}$$

where

$$c_i = a_i^w + b_i^w\ (i = 1, \ldots, M). \tag{45}$$

**Statement 14.** Problem $\langle a, b, S\rangle \in P(S)$ (i.e. proportional divisions exist) if and only if at least one of inequalities systems (42), (43), (45) is compatible.

**Proof.** Assume

$$T_1(\sigma) = \sum_{i=1}^{M} \sigma_i a_i^w,\ T_2(\sigma) = \sum_{i=1}^{M} (1 - \sigma_i) b_i^w,\ T(\sigma) = (T_1(\sigma), T_2(\sigma)). \tag{46}$$

Let us consider set $A(\sigma) = T(\sigma) + A_d(a, b, S)$ (see formula (10)). $A(\sigma)$ is a shifted polygon $A_d(a, b, S)$. By definition, proportional division exist if and only if polygon $A(\sigma)$ intersects set $P$ of the plane, defined as follows:

$$P = \{(x, y)\mid x \geq H/2, y \geq H/2\}$$

at least for one Boolean vector $\sigma$.
It can happen in one of the three cases:
1) left vertex of $A(\sigma)$ (point $(T_1(\sigma), T_2(\sigma) + S_2)$ belongs to $P$;
2) right vertex of $A(\sigma)$ (point $(T_1(\sigma) + S_1, T_2(\sigma))$ belongs to $P$;
3) vertices defined in 1) and 2) do not belong to $P$, but at least one point of $A(\sigma)$ belongs to $P$.

In the 1st case we have inequalities (see (41) and (46)):

$$\sum_{i=1}^{M} \sigma_i a_i^w \geq H/2, \tag{47a}$$

$$\sum_{i=1}^{M} (1 - \sigma_i) b_i^w + \sum_{i=1}^{L} b_i^d \geq H/2. \tag{47b}$$

Rearranging left-hand side of (47b), we have

$$\sum_{i=1}^{M} (1 - \sigma_i) b_i^w + \sum_{i=1}^{L} b_i^d = \sum_{i=1}^{M} b_i^w - \sum_{i=1}^{M} \sigma_i b_i^w + \sum_{i=1}^{L} b_i^d = \sum_{i=1}^{M} b_i^w + \sum_{i=1}^{L} b_i^d - \sum_{i=1}^{M} \sigma_i b_i^w = H - \sum_{i=1}^{M} \sigma_i b_i^w \geq H/2$$ that implies

$$\sum_{i=1}^{M} \sigma_i b_i^w \leq H/2. \tag{47c}$$

(47a) and (47c) together coincide with system (43).
Analogously, in the 2nd case we receive system (42).
In the 3rd case simple geometrical reasoning shows that the segment connecting points $(T_1(\sigma) + S_1, T_2(\sigma))$ and $(T_1(\sigma), T_2(\sigma) + S_2)$ must intersect the rayon $x = y$. In the intersection point we receive that for some $\lambda \in [0, 1]$

$$\sum_{i=1}^{M} \sigma_i a_i^w + (1-\lambda)S_1 = \sum_{i=1}^{M} (1 - \sigma_i) b_i^w + \lambda S_2.$$

Expressing $\lambda$ from this equation and taking into account inequalities $\lambda \geq 0$, $\lambda \leq 1$, after rearranging receive system (44).

**Statement 15.** Problem $\langle a, b, S\rangle \in Q(S)$ (i.e. an equitable division exists) if and only if inequalities system (44) is compatible.

**Proof.** See the 3rd case in the previous proof.

To be useful, statements 14 and 15 must be completed with some efficient algorithms of compatibility checking for systems (42), (43) and (44). In order to find them consider the following optimization problems relatively Boolean variables $\sigma_1, \ldots, \sigma_N$:

$$\sum_{i=1}^{N} \sigma_i b_i \to \max \tag{48a}$$

subject to

$$\sum_{i=1}^{N} \sigma_i a_i \leq H/2; \tag{48b}$$

$$\sum_{i=1}^{N} \sigma_i a_i \to \max \tag{49a}$$

subject to

$$\sum_{i=1}^{N} \sigma_i b_i \leq H/2; \tag{49b}$$

$$\sum_{i=1}^{N} \sigma_i c_i \to \max \tag{50a}$$

subject to

$$\sum_{i=1}^{N} \sigma_i c_i \leq H. \tag{50b}$$

The following statement is evident.

**Statement 16.**
System (42) is compatible if and only if the maximal value in problem (48) is not less than $H/2$.
System (43) is compatible if and only if the maximal value in problem (49) is not less than $H/2$.
System (44) is compatible if and only if the maximal value in problem (50) is not less than $H - (S_1 + S_2)$.

Thus, statements 14 – 16 reduce checking existence of proportional and equitable division problem to optimization problems relatively Boolean variables $\sigma_1, \ldots, \sigma_N$. Namely:

Problem (48) is a knapsack problem with item weights $b_1, \ldots, b_N$, item values $a_1, \ldots, a_N$ and weight restriction $H/2$.

Problem (49) is a knapsack problem with item weights $a_1, \ldots, a_N$, item values $b_1, \ldots, b_N$ and weight restriction $H/2$.

Problem (50) is a simplest knapsack problem, where values $c_1, \ldots, c_N$ coincide with weights and weight restriction is $H$.

All the above mentioned knapsack problems can be efficiently solved by dynamic programming (see detail in section 4.1).

## 6. Examples

In this section we give some relatively simple examples that illustrate the results of sections 5 and 6.

**Example 4.** Let us consider the following division problem with one divisible and four indivisible items (divisible items here and in next examples are at the beginning):

Table 8

| A | B |
|---|---|
| 1 | 1 |
| 45 | 30 |
| 30 | 25 |
| 15 | 22 |
| 9 | 22 |
| 100 | 100 |

For this problem:
system (42) is

$$\begin{cases} 45\sigma_1 + 30\sigma_2 + 15\sigma_3 + 9\sigma_4 \leq 50 \\ 30\sigma_1 + 25\sigma_2 + 22\sigma_3 + 22\sigma_4 \geq 50 \end{cases} \quad (51)$$

system (43) is

$$\begin{cases} 45\sigma_1 + 30\sigma_2 + 15\sigma_3 + 9\sigma_4 \geq 50 \\ 30\sigma_1 + 25\sigma_2 + 22\sigma_3 + 22\sigma_4 \leq 50 \end{cases} \quad (52)$$

$c_1 = 45+30 = 75$, $c_2 = 30+25 = 55$, $c_3 = 15+22 = 37$, $c_3 = 9+22 = 31$;
system (44) is

$$\begin{cases} 75\sigma_1 + 55\sigma_2 + 37\sigma_3 + 31\sigma_4 \leq 100 \\ 75\sigma_1 + 55\sigma_2 + 37\sigma_3 + 31\sigma_4 \geq 98 \end{cases} \quad (53)$$

In order to check compatibility of these three systems, consider the set of all the 16 Boolean vectors $(\sigma_1,\sigma_2,\sigma_3,\sigma_4)$ and calculate the corresponding value of three expressions:

$$V_1 = 45\sigma_1 + 30\sigma_2 + 15\sigma_3 + 9\sigma_4,$$

$$V_2 = 30\sigma_1 + 25\sigma_2 + 22\sigma_3 + 22\sigma_4,$$

$$V_3 = 75\sigma_1 + 55\sigma_2 + 37\sigma_3 + 31\sigma_4 \ (V_3 = V_1 + V_2).$$

Results of these simple calculations are presented in Table 9:

Table 9

| $\sigma_1$ | $\sigma_2$ | $\sigma_3$ | $\sigma_4$ | $V_1$ | $V_2$ | $V_3$ | | $\sigma_1$ | $\sigma_2$ | $\sigma_3$ | $\sigma_4$ | $V_1$ | $V_2$ | $V_3$ |
|---|---|---|---|---|---|---|---|---|---|---|---|---|---|---|
| 0 | 0 | 0 | 0 | 0 | 0 | 0 | | 1 | 0 | 0 | 0 | 45 | 30 | 75 |
| 0 | 0 | 0 | 1 | 9 | 22 | 31 | | 1 | 0 | 0 | 1 | 54 | 52 | 102 |
| 0 | 0 | 1 | 0 | 15 | 22 | 37 | | 1 | 0 | 1 | 0 | 60 | 52 | 112 |
| 0 | 0 | 1 | 1 | 24 | 44 | 68 | | 1 | 0 | 1 | 1 | 69 | 74 | 143 |
| 0 | 1 | 0 | 0 | 30 | 25 | 55 | | 1 | 1 | 0 | 0 | 75 | 55 | 130 |
| 0 | 1 | 0 | 1 | 39 | 47 | 86 | | 1 | 1 | 0 | 1 | 84 | 77 | 161 |
| 0 | 1 | 1 | 0 | 45 | 47 | 92 | | 1 | 1 | 1 | 0 | 90 | 77 | 167 |
| 0 | 1 | 1 | 1 | 54 | 69 | 123 | | 1 | 1 | 1 | 1 | 99 | 99 | 198 |

The compatibility of system (51) means that in some row of the table $V_1 \leq 50$, $V_2 \geq 50$; the compatibility of system (52) means that in some row of the table $V_1 \geq 50$, $V_2 \leq 50$; the compatibility of system (53) means that in some row of the table $98 \leq V_3 \leq 100$. None of rows satisfies these conditions. Thus, in the considered case proportional division does not exist, that can be written in the form

$$P(S) \subset E(S) \quad (54)$$

at least for $S = \langle 1, 4, 100 \rangle$.

**Example 5.** Let us consider the following division problem with one divisible and four indivisible items:

Table 10

| A | B |
|---|---|
| 3 | 3 |
| 45 | 17 |
| 30 | 20 |
| 20 | 22 |
| 2 | 38 |
| 100 | 100 |

In this case formulas (6) for gains are written as follows:

$$G_A(x, \sigma) = xa_0 + \sum_{i=1}^{N}\sigma_i a_i = 3x+45\sigma_1+30\sigma_2+20\sigma_3+2\sigma_4, \quad (55)$$

$$G_B(x, \sigma) = (1-x)b_0 + \sum_{i=1}^{N}(1-\sigma_i)b_i = -xb_0 = 3(1-x) +17(1-\sigma_1) +20(1-\sigma_2)+22(1-\sigma_3)+ 38(1-\sigma_4). \quad (56)$$

From (55) and (56) immediately

$$G_A(x, \sigma) = 3x + V_1,$$
$$G_B(x, \sigma) = 3(1-x) + V_2,$$
$$G_A(x, \sigma) - G_B(x, \sigma) = (6x-3) + (V_1 - V_2),$$

where

$$V_1 = 45\sigma_1+30\sigma_2+20\sigma_3+2\sigma_4, \quad V_2 = 17(1-\sigma_1) +20(1-\sigma_2)+22(1-\sigma_3)+ 38(1-\sigma_4).$$

Because $0 \leq x \leq 1$, it is clear that existence of an equitable division implies that for some Boolean vector $\sigma$

$$|V_1 - V_2| \leq 3.$$

Assume $V_3 = 62\sigma_1+50\sigma_2+42\sigma_3+40\sigma_4$; system (44) in the considered case becomes

$$94 \leq V_3 \leq 100.$$

Let us consider the set of all the 16 Boolean vectors $(\sigma_1,\sigma_2,\sigma_3,\sigma_4)$ and calculate the corresponding value of $V_1$, $V_2$ and $V_3$:

Table 11

| $\sigma_1$ | $\sigma_2$ | $\sigma_3$ | $\sigma_4$ | $V_1$ | $V_2$ | $V_3$ |
|---|---|---|---|---|---|---|
| 0 | 0 | 0 | 0 | 0 | 97 | 0 |
| 0 | 0 | 0 | 1 | 2 | 59 | 40 |
| 0 | 0 | 1 | 0 | 20 | 75 | 42 |
| 0 | 0 | 1 | 1 | 22 | 37 | 82 |
| 0 | 1 | 0 | 0 | 30 | 77 | 50 |
| 0 | 1 | 0 | 1 | 32 | 39 | 90 |
| 0 | 1 | 1 | 0 | 50 | 55 | 92 |
| 0 | 1 | 1 | 1 | 52 | 17 | 132 |

| $\sigma_1$ | $\sigma_2$ | $\sigma_3$ | $\sigma_4$ | $V_1$ | $V_2$ | $V_3$ |
|---|---|---|---|---|---|---|
| 1 | 0 | 0 | 0 | 45 | 80 | 62 |
| 1 | 0 | 0 | 1 | 47 | 42 | 102 |
| 1 | 0 | 1 | 0 | 65 | 58 | 104 |
| 1 | 0 | 1 | 1 | 67 | 20 | 144 |
| 1 | 1 | 0 | 0 | 75 | 60 | 112 |
| 1 | 1 | 0 | 1 | 77 | 22 | 152 |
| 1 | 1 | 1 | 0 | 95 | 38 | 154 |
| 1 | 1 | 1 | 1 | 97 | 0 | 184 |

The compatibility of system (44) means that in some row of the table $94 \leq V_3 \leq 100$. None of rows satisfies this condition. Thus, by statement 11 in the considered case an equitable division does not exist. Indeed, for any row $|V_1 - V_2| > 3$, and this difference cannot be compensate be one divisible item with value 3 for both participants. At the same time in several rows of the table (namely, 0110, 1010, 1100) $V_1 \geq 50$, $V_2 \geq 50$ that means that proportional divisions exist. This example implies inclusion

$$Q(S) \subset P(S) \quad (57)$$

at least for S =⟨1, 4, 100⟩.

**Example 6.** Let us consider the following division problem with one divisible and four indivisible items:

Table 12

| A | B |
|---|---|
| 5 | 5 |
| 40 | 49 |
| 10 | 1 |
| 20 | 25 |
| 25 | 20 |
| 100 | 100 |

In this case system we easily find equitable divisions

$$\{1, 2\}, \{0, 3, 4\} \text{ and } \{0, 3, 4\}, \{1, 2\}.$$

In both cases each participant receives exactly 50; one of these divisions is a complement division for another. But this division is not efficient because division corresponding $\sigma = (0,1,1,1)$: division {2, 3,

4} for A and {0, 1} for B gives more for both participants: (55, 54). Thus, this example implies inclusion

$$F(S) \subset Q(S) \tag{58}$$

at least for at least for S =⟨1, 4, 100⟩.
Summing inclusion (54), (57) and (58), we have

$$F(S) \subset Q(S) \subset P(S) \subset E(S) = U(S),$$

in opposite to the case in which all items are divisible, where all the inclusions are equalities (see statement 2).

Note, that the same inclusion is true for the problem from Example 3 with the other signature S = ⟨2, 3, 100⟩.

**Example 7\*.** Let us consider the following division problem with one divisible and four indivisible items:

Table 13

| A | B |
|---|---|
| 17 | 17 |
| 42 | 45 |
| 37 | 34 |
| 2 | 2 |
| 2 | 2 |
| 100 | 100 |

For this problem:
system (42) is

$$\begin{cases} 42\sigma_1 + 37\sigma_2 + 2\sigma_3 + 2\sigma_4 \leq 50 \\ 45\sigma_1 + 34\sigma_2 + 2\sigma_3 + 2\sigma_4 \geq 50 \end{cases} \tag{59}$$

system (43) is

$$\begin{cases} 42\sigma_1 + 37\sigma_2 + 2\sigma_3 + 2\sigma_4 \geq 50 \\ 45\sigma_1 + 34\sigma_2 + 2\sigma_3 + 2\sigma_4 \leq 50 \end{cases} \tag{60}$$

$c_1 = 42+45 = 87$, $c_2 = 37+34 = 71$, $c_3 = 2 + 2 = 4$, $c_4 = 2+2 = 41$;

system (44) is

$$\begin{cases} 87\sigma_1 + 71\sigma_2 + 4\sigma_3 + 4\sigma_4 \leq 100 \\ 87\sigma_1 + 71\sigma_2 + 4\sigma_3 + 4\sigma_4 \geq 66 \end{cases} \tag{61}$$

In order to check compatibility of these three systems, consider the set of all the 16 Boolean vectors $(\sigma_1, \sigma_2, \sigma_3, \sigma_4)$ and calculate the corresponding value of three expressions:

$$V_1 = 42\sigma_1 + 37\sigma_2 + 2\sigma_3 + 2\sigma_4,$$
$$V_2 = 45\sigma_1 + 34\sigma_2 + 2\sigma_3 + 2\sigma_4,$$
$$V_3 = 87\sigma_1 + 71\sigma_2 + 4\sigma_3 + 4\sigma_4 \ (V_3 = V_1 + V_2).$$

Results of these simple calculations are presented in Table14:

Table 14

| $\sigma_1$ | $\sigma_2$ | $\sigma_3$ | $\sigma_4$ | $V_1$ | $V_2$ | $V_3$ | $\sigma_1$ | $\sigma_2$ | $\sigma_3$ | $\sigma_4$ | $V_1$ | $V_2$ | $V_3$ |
|---|---|---|---|---|---|---|---|---|---|---|---|---|---|
| 0 | 0 | 0 | 0 | 0 | 0 | 0 | 1 | 0 | 0 | 0 | 42 | 45 | 87 |
| 0 | 0 | 0 | 1 | 2 | 2 | 4 | 1 | 0 | 0 | 1 | 44 | 47 | 91 |
| 0 | 0 | 1 | 0 | 2 | 2 | 4 | 1 | 0 | 1 | 0 | 44 | 47 | 91 |
| 0 | 0 | 1 | 1 | 4 | 4 | 8 | 1 | 0 | 1 | 1 | 46 | 49 | 95 |
| 0 | 1 | 0 | 0 | 37 | 34 | 71 | 1 | 1 | 0 | 0 | 79 | 79 | 158 |
| 0 | 1 | 0 | 1 | 39 | 36 | 75 | 1 | 1 | 0 | 1 | 81 | 81 | 162 |
| 0 | 1 | 1 | 0 | 39 | 36 | 75 | 1 | 1 | 1 | 0 | 81 | 81 | 162 |
| 0 | 1 | 1 | 1 | 41 | 38 | 79 | 1 | 1 | 1 | 1 | 83 | 83 | 166 |

\*This example was suggested by SU – HSE student Alexander Shalenny

The compatibility of system (59) means that in some row of the table $V_1 \leq 50$, $V_2 \geq 50$; the compatibility of system (60) means that in some row of the table $V_1 \geq 50$, $V_2 \leq 50$; the compatibility of system (61) means that in some row of the table $66 \leq V_3 \leq 100$. None of rows satisfies first two conditions, though there are 8 rows satisfying (61). Thus, in the considered case equitably fair division exists: for instance, participant A receives item 3 (37%) and 14,5 of item 1 (14,5%), while participant B receives items 2, 4, 5 (49%) and 2,5 of item 1 (2,5%). Thus, everyone receives 51,5%; no division gives more for both. Therefore, this is a fair division (it is simultaneously profitably, uniformly and equitably fair).

**Example 8.** Let us consider the following division problem with three indivisible items:

Table 15

| A | B |
|---|---|
| 51 | 40 |
| 45 | 50 |
| 4 | 10 |
| 100 | 100 |

In this problem point $(G_A(\sigma), (G_B(\sigma))$ is the pair of gains received by participants as a result of division $\sigma$, where

$$G_A(\sigma) = 51\sigma_1 + 45\sigma_2 + 4\sigma_3,$$

$$G_B(\sigma) = 40(1-\sigma_1) + 50(1-\sigma_2) + 10(1-\sigma_3).$$

Let us calculate gains $G_A(\sigma)$, $G_B(\sigma)$ and expressions

$$\min\{G_A(\sigma), G_B(\sigma)\}, \tag{62}$$

$$|G_A(\sigma) - G_B(\sigma)| \tag{63}$$

for all 8 Boolean vectors $(\sigma_1, \sigma_2, \sigma_3)$ (see statement 7 and steps 11, 12 of the essential algorithm). The results of calculation are presented in Table 16:

Table 16

| $\sigma_1$ | $\sigma_2$ | $\sigma_3$ | $G_A(\sigma)$ | $G_B(\sigma)$ | $\min\{G_A(\sigma), G_B(\sigma)\}$ | $|G_A(\sigma) - G_B(\sigma)|$ |
|---|---|---|---|---|---|---|
| 0 | 0 | 0 | 0 | 100 | 0 | 100 |
| 0 | 0 | 1 | 4 | 90 | 4 | 86 |
| 0 | 1 | 0 | 45 | 50 | 45 | 5 |
| 0 | 1 | 1 | 49 | 40 | 40 | 9 |
| 1 | 0 | 0 | 51 | 60 | 51 | 9 |
| 1 | 0 | 1 | 55 | 50 | 50 | 5 |
| 1 | 1 | 0 | 96 | 10 | 10 | 86 |
| 1 | 1 | 1 | 100 | 0 | 0 | 100 |

Table 16 clear demonstrates that criteria (62) and (63) defining profitable and uniform fairness lead to different divisions. There are only two rows (5[th] and 6[th]) presenting proportional divisions. One of them maximizes (62) while the other minimizes (63) over the set of all the efficient divisions.

## 7. Manipulation in the divisible case

Usually, presentation of false data – false importance values in this case – in order to receive more is called manipulation. The well known manipulation problem is considered in the following game form.
1. Players independently choose arbitrary integer coefficients $a_1, \ldots, a_N$ and $b_1, \ldots, b_N$ that are consider as their values of items.
2. A fair division $\langle x, \sigma \rangle$ is constructed based on the presented values $a = (a_1, \ldots, a_N)$ and $b = (b_1, \ldots, b_N)$.
3. The gain of player A is equal to value of division $\langle x, \sigma \rangle$ from the point of view of his true valuations $a_i^t, \ldots, a_N^t$ (analogously for player B). Formally

$$G_A(a, b) = \sum_{i=1}^{M} \sigma_i a_i^t + \sum_{i=1}^{L} x_i a_i^t, \tag{64a}$$

$$G_B(a, b) = \sum_{i=1}^{M} (1 - \sigma_i) b_i^t + \sum_{i=1}^{L} (1 - x_i) b_i^t, \tag{64b}$$

where $\langle x, \sigma \rangle$ is a fair division corresponding to the presented values $a$ and $b$ (not to the true values $a^t$ and $b^t$).

Thus, the game of two persons with finite number of strategies is defined. Remember, that strategy $a^*$ of player A is a guarantying strategy, if the minimal gain of this strategy is the maximal one, or

$$\min_b G_A(a^*, b) \geq \min_b G_A(a, b)$$

(analogously for player B:

$$\min_a G_B(a, b^*) \geq \min_a G_B(a, b)).$$

If guarantying strategy $a^*$ differs from true strategy $a^t$, that means that manipulation (i.e. presentation of strategy $a^*$ instead true strategy $a^t$) is profitable for player A; otherwise, manipulation can decrease the gain of the player (the same for player B).

**Statement 17.** In the considered case $a^* = a^t$, $b^* = b^t$, i.e. manipulation can decrease the gains.

**Proof.** Suppose that $a^* \neq a^t$. If player B chooses strategy $a^t$ of player A, then player B receives more than a half <u>in values $a^t$</u> (see Statement 2). Hence, player A receives less than a half in <u>his own values $a^t$</u> (see (64)). At the same time, if player A chooses his true strategy $a^t$, then he receives at least half in <u>his own values $a^t$</u> independently of any choice of B. Therefore, strategy $a^t$, is the guarantying strategy of A (analogously for B).

**8. Conclusion**

In the connection with the presented material two typical questions arise:
- How to modified the results to the case of $N$ participants, where $N > 2$?
- How to cope with manipulation in general case?

These questions, as well as some others, are out of scope of this material. However, the author intends to continue investigation in this wide domain. Particularly, it seems of expedient to consider division problems in which some groups of items may be of special interest for participants, so that value of a group significantly exceeds the sum of values of its separate items (package deals).

The author is grateful to his colleagues F.T. Aleskerov and D.A. Shwarts for their support and attention to this work.